\documentclass[ps,prd,preprintnumbers,superscriptaddress,nofootinbib,floatfix,twocolumn,notitlepage]{revtex4-2}
\pdfoutput=1 

\usepackage{dcolumn}

\usepackage{booktabs}
\usepackage{tabularx}

\usepackage{mathrsfs}
\usepackage{bm,amssymb,slashed,graphicx,multirow,soul,mathtools,xspace,array,tikz,amsmath,siunitx}
\usepackage[compat=1.1.0]{tikz-feynman}   
\usepackage{float}                   
\allowdisplaybreaks
\usepackage{ bbold }
\usepackage{caption,subcaption}
\usepackage{braket}
\usepackage{hyperref}
\usepackage{colortbl}

\definecolor{nicered}{rgb}{0.7,0.1,0.1}
\definecolor{nicegreen}{rgb}{0.1,0.5,0.1}
\definecolor{violet}{rgb}{0.7,0.3,0.3}
\hypersetup{colorlinks,citecolor= nicegreen,linkcolor= nicered}

\newcommand{\lp}{\left(}
\newcommand{\rp}{\right)}

\newcommand{\g}{\gamma}

\newcommand{\beq}{\begin{equation} }
\newcommand{\eeq}{\end{equation}} 
\newcommand{\bi}{\begin{itemize} }
\newcommand{\ei}{\end{itemize} }

\definecolor{Red}{rgb}{1.,0.,0.}
\definecolor{Grn}{rgb}{0.,0.75,0.}
\definecolor{Blu}{rgb}{0.,0.,1.}
\definecolor{Pink}{rgb}{1,0.08,0.58}

\let\Re\relax
\DeclareMathOperator{\Re}{Re}
\let\Im\relax
\DeclareMathOperator{\Im}{Im}

\usepackage[T1]{fontenc} 

\usepackage{amsmath,amssymb,epsfig,ulem,color,slashed}
\allowdisplaybreaks  

\newcommand{\at}{A_{\rm T}^{\mu3e}}
\newcommand{\Op}{\mathcal{O}}
\newcommand{\BR}{\mathrm{BR}}
\newcommand{\JH}{i H^\dag \overleftrightarrow{D_\mu} H}
\newcommand{\JHa}{i H^\dag \overleftrightarrow{D_\mu^a} H}

\newcommand{\CC}{\mathcal{C}}

\begin{document} 


\title{\boldmath Large CP violation in flavor violating muon decays}


\author{Diego Redigolo}
\email{diego.redigolo@fi.infn.it}
\author{Michele Tammaro}
\email{michele.tammaro@fi.infn.it}
\author{Andrea Tesi}
\email{andrea.tesi@fi.infn.it}
\affiliation{INFN Sezione di Firenze, Via G. Sansone 1, I-50019 Sesto Fiorentino, Italy
}

\date{\today}

\begin{abstract}
We identify new room for CP violation in lepton flavor violating observables not bound by the electric dipole moment of leptons. By focusing on new physics in the muon-electron sector, we show that CP violation can make its first appearance in lepton flavor violating muon decays rather than in the electric dipole moment of the electron, further motivating the experimental program of Mu3e. We tackle this issue by performing the full one-loop running and matching from the low energy observables at the muon scale, including the T-odd asymmetry in $\mu\to3e$ decays, to the Standard Model effective field theory above the electroweak scale. We then sketch a simple UV model that can give rise to these patterns. 
\end{abstract}

\maketitle

%
\section{Introduction}
\label{sec:intro}
%
One of the most precise measurement in particle physics is the null-result of the electric dipole moment of the electron (eEDM), $d_e$. It is found that $|d_e|\leq10^{-29}e$\,cm~\cite{ACME:2018yjb}, with future experiments projected to probe $|d_e|\leq10^{-31}e$\,cm~\cite{Meisenhelder:2023qcq}. This measurement alone can probe new physics injecting a new source of CP Violation (CPV) up to scales of ${\cal O}(10)$ TeV~\cite{Panico:2018hal,Cesarotti:2018huy} even if the new physics respects the flavour structure of the SM. 

It is then bizarre to imagine that large CPV effects can be visible in the lepton sector given the current bounds on the $d_e$. In this paper we explore this possibility, leveraging on the large improvements expected in probing Lepton Flavor Violating (LFV) observables in the near future. In particular, we focus our attention on Mu3e~\cite{Mu3e:2020gyw}, which will reach an unprecedented sensitivity on the rate of $\mu\to3e$ and offers a unique possibility to test CP-violation as first discussed in Ref.~\cite{Okada:1999zk,Kuno:1999jp, Bolton:2022lrg} and summarized in Sec.~\ref{sec:Mu3e:AT}.

The main goal of this letter is to look for models that maximize the effect of CP violation in LFV operators, while still being consistent with the eEDM limits. Within an effective field theory approach, we first quantify the flavor alignment needed for LFV muon decays to be consistent with the eEDM. This is easily done in Sec.~\ref{sec:LEFT} in the context of the Low Energy Effective Field Theory (LEFT) of muons, electrons and photons which is the relevant one for the EDMs and LFV observables. In Sec.~\ref{sec:Operators} we then map the required alignment to possible UV completions, by matching the LEFT with the Standard Model Effective Field Theory (SMEFT). The latter will serve as computational framework to include running effects from large to low scales. Finally, in Sec.~\ref{sec:UVmodels} we explore possible UV realizations and draw our conclusions in Sec.~\ref{sec:Conclusions}.

%

\begin{table}[t]
    \caption{\label{tab:bounds}Lepton observables considered in this work. In the second column we show the present bounds and in the third column the projected sensitivity of the forthcoming next generation experiments: ACME~III for electron EDM, muEDM for muon EDM, MEG II for $\mu\to e \gamma$, Mu3e for $\mu\to3e$ and Mu2e and COMET for $\mu N\to e N$ conversion.}

    \renewcommand{\arraystretch}{1.3}
    \begin{tabular}{c|r|r}
        Observables & Current bound & Future sensitivity \\ \hline
        $|d_e|/e $ & $1.1\times10^{-29}~{\rm cm}$~\cite{ACME:2018yjb} & $10^{-31}~{\rm cm}$~\cite{Meisenhelder:2023qcq} \\
         $|d_\mu|/e $ & $1.8\times 10^{-19}~{\rm cm}$~\cite{Muong-2:2008ebm} & $6\times 10^{-23}$\text{  cm}~\cite{Adelmann:2021udj} \\
        ${\rm Br}(\mu\to e\g) $ & $3.1\times10^{-13}$~\cite{MEG:2016leq,MEGII:2023ltw} & $6\times10^{-14}$~\cite{MEGII:2018kmf} \\
        ${\rm Br}(\mu\to 3e) $ & $10^{-12}$~\cite{BELLGARDT19881} & $5\times10^{-16}$~\cite{Blondel:2013ia, Mu3e:2020gyw}\\
        ${\rm CR}(\mu N\to e N) $ & $7\times10^{-13}$~\cite{SINDRUMII:2006dvw} & $10^{-16}$~\cite{Mu2e:2022ggl,COMET:2018auw} \\
    \end{tabular}
\end{table}

%

%
\section{LEFT framework}
\label{sec:LEFT}
%

The set of muon decay observables that we consider is summarized in Table~\ref{tab:bounds}, where we also report the respective current and future limits. These are: the LFV radiative decays, $\mu\to e\g$; the $\mu\to 3e$ branching ratio; the $\mu-e$ conversion rate in nuclei (see Ref.~\cite{Calibbi:2017uvl} for a review). These observables complement the electron EDM present and future measurements. 

In order to parametrize new LFV phyiscs in these channels, we consider an effective lagrangian valid at the muon mass scale;
at these energies the EFT only contains electrons, muons and photons. The relevant operators are 
\begin{equation}\label{eq:LEFT}
\begin{split}\
\mathscr{L}
&=\frac{c_V^{LL}}{\Lambda^2}\bar{\mu}_L\gamma^\mu e_L \bar{e}_L\gamma_\mu e_L 
+\frac{c_V^{LR}}{\Lambda^2}\bar{\mu}_L\gamma^\mu e_L \bar{e}_R\gamma_\mu e_R\\
&+\frac{c_V^{RL}}{\Lambda^2}\bar{\mu}_R\gamma^\mu e_R \bar{e}_L\gamma_\mu e_L
+\frac{c_V^{RR}}{\Lambda^2}\bar{\mu}_R\gamma^\mu e_R \bar{e}_R\gamma_\mu e_R \\
&+ \frac{c_{\ell\g}^{\mu e} m_\mu}{\Lambda^2} \bar{\mu}_L \sigma_{\mu\nu} e_R F^{\mu\nu} + \frac{c_{\ell\g}^{e \mu} m_\mu}{\Lambda^2} \bar{e}_L \sigma_{\mu\nu} \mu_R F^{\mu\nu} \\
&+\frac{c_{\ell\g}^{\mu \mu} m_\mu}{\Lambda^2} \bar{\mu}_L \sigma_{\mu\nu} \mu_R F^{\mu\nu} + \frac{c_{\ell\g}^{e e} m_\mu}{\Lambda^2} \bar{e}_L \sigma_{\mu\nu} e_R F^{\mu\nu} \\
&+ \text{ h.c.} \,,
\end{split}
\end{equation}
where $\Lambda$ is the typical UV energy scale, while $c$ are dimensionless Wilson coefficients. We did not include scalar 4-lepton operators which will not be generated in SMEFT at dimension six~\cite{Jenkins:2017jig}. Note that neither neutrinos nor nuclei appear in this basis of operators. Both kind of operators will give a largely subleading contributions to respective observables, as neutron EDM or electroweak precision measurements. On a different footing, we include $\tau$ leptons contributions as these can give sizeble threshold corrections when integrated out. The full list of LEFT operators can be found in Refs.~\cite{Jenkins:2017jig, Dekens_2019}.

Lastly, note that the relevant scale for the eEDM is the electron mass, $m_e$, thus the relevant LEFT is obtained by integrating out the muon from Eq.~\eqref{eq:LEFT}. In particular, the Wilson coefficients need to be evaluated at the $m_e$ scale. However, the running is dominated by QED effects and the threshold matching is proportional to $m_\mu$, thus we expect these corrections to be negligible. That is that, except from the $\tau$ threshold matching mentioned above, we can consider the LEFT coefficients to be unaffected by the low energy running and take $c(m_\mu) = c(\mu_W)$, where $\mu_W\sim160$ GeV is the electroweak symmetry breaking scale.

%
\section{Observables at the $\mu$ mass threshold}
\label{sec:Observables}
%

Here we review the low energy observables and show the contribution they receive from the operators in Eq.~\eqref{eq:LEFT}.

\paragraph*{\bf Lepton electric dipole moments:}
the electric dipole moment can be directly extracted from the flavor diagonal entries of the ${\cal O}_{f\g}$ operator. We have 
\beq\label{eq:moments}
d_\ell = -\frac{2 m_\ell}{\Lambda^2} \Im[c_{\ell\g}^{\ell\ell}(m_{\ell})]\,.
\eeq
Unlike other observables in this Section, a non-zero EDM is a direct probe of CPV effects. The strongest bounds have been obtained by the ACME experiment~\cite{ACME:2018yjb}, as $|d_e|\leq10^{-29}e$\,cm, with a two order of magnitude improvement expected at ACME III~\cite{Meisenhelder:2023qcq}.

Given that the naive scaling of the muon EDM $d_\mu\sim m_\mu/m_e d_e$ is respected in our setup, even though the sensitivity on the muon EDM  will be impressively improved in the near future~\cite{Adelmann:2021udj} the current electron EDM bound will still outperform it by several orders of magnitude.  

\paragraph*{\bf Radiative lepton decays:} 
the flavor off-diagonal entries of the dipole operator will induce radiative decays of leptons into lighter states. Focusing on the muon, such operators generate the $\mu\to e\g$ decay whose branching ratio reads  
\beq\label{eq:radiative}
\BR(\mu \to e \g) = (|c_{\ell\g}^{\mu e}|^2+ |c_{\ell\g}^{e\mu}|^2)\frac{m_{\mu}^5}{4\pi\Lambda^4\Gamma_\mu} \,,
\eeq
where $\Gamma_\mu\simeq G_F^2m_\mu^5/(192\pi^3)=3\times 10^{-10}\text{ eV}$ is the muon width and the coefficients are taken at the muon mass threshold.
The current combination of the MEG~\cite{MEG:2016leq} and MEG~II~\cite{MEGII:2023ltw} measurements constrains the branching ratio to be ${\rm Br}(\mu\to e\gamma)<3.1\times10^{-13}$ with MEGII~\cite{MEGII:2018kmf} prospected to reach ${\rm Br}(\mu\to e\gamma)<6\times10^{-14}$.

\paragraph*{\bf $\mu\to 3 e$ decay:} the muon decay in 3 electrons can be mediated again by the dipole operator, or directly via the four-fermion operators $\Op^V$. The full decay width can be found in Refs.~\cite{Okada:1999zk,Kuno:1999jp, Bolton:2022lrg}, which we use in our numerical analysis. In the limit where only $c_{\ell\gamma}$ dominates, the $\mu\to3e$ branching ratio can be written
\begin{equation}\label{eq:mu3e}
\BR(\mu \to 3e)\approx \frac{\alpha}{3\pi} \lp \ln\frac{m_\mu^2}{m_e^2} - 3 \rp \times \BR(\mu\to e\g)\,.
\end{equation}
and it is directly proportional to the $\mu\to e\gamma$ branching ratio in Eq.~\eqref{eq:radiative}. The current bound was set by SINDRUM in 1988~\cite{BELLGARDT19881}, as ${\rm Br}(\mu\to3e)\leq10^{-12}$. The proposed Mu3e experiment~\cite{Blondel:2013ia, Mu3e:2020gyw} will substantially improve this limit to ${\rm Br}(\mu\to3e)\leq2\times10^{-15}$ in phase-I and ${\rm Br}(\mu\to3e)\leq5\times10^{-16}$ in phase-II. We will use the latter expected sensitivity in our analysis.

\paragraph*{\bf $\mu\to e$ conversion:} muons can convert to electrons in the electromagnetic field of a nucleus $N$, if the off-diagonal entry of the dipole operator is non-zero. The conversion rate (CR) can be written as
\beq
{\rm CR}(\mu N\to eN) = \frac{\Gamma(\mu N\to e N)}{\Gamma_{\rm capt.}(\mu N)}\,,
\eeq
where $\Gamma_{\rm capt.}(\mu N)$ is the muon capture rate of the nucleus~\cite{Kitano:2002mt}. Proper predictions of the conversion rate can be obtained by matching the full LEFT lagrangian into nuclear matrix elements~\cite{Haxton:2024lyc}; approximately, in our LEFT we can write
\beq
{\rm CR}(\mu N\to e N) \approx \alpha \times {\cal B}(\mu\to e\g)\,,
\eeq
where for this scaling to be true it is instrumental that no operators with quarks are present in Eq.~\eqref{eq:LEFT}. As we will discuss in more detail in Sec.~\ref{sec:Operators}, the presence of four-fermion operators with quarks would enhance the $\mu\to e$ conversion rate to a level that will make it impossible to observe CPV in $\mu\to 3 e$.

The strongest limit on $\mu\to e$ conversion so far has been obtained by SINDRUM II using Gold~\cite{SINDRUMII:2006dvw}, which gives ${\rm CR}(\mu\,{\rm Au}\to e\,{\rm Au}) < 7.0\times10^{-13}$, and Titanium~\cite{Wintz:1998rp}, which requires ${\rm CR}(\mu~{\rm Ti}\to e~{\rm Ti}) < 6.1\times10^{-13}$. The Mu2e Collaboration~\cite{Mu2e:2014fns}, using Aluminium, projects to reach a far stronger limit, ${\rm CR}(\mu~{\rm Al}\to e~{\rm Al}) < 10^{-16}$~\cite{Mu2e:2022ggl}. The COMET experiment~\cite{COMET:2018auw} prospects to reach ${\rm CR}(\mu\,{\rm Al}\to e\,{\rm Al}) < 7\times10^{-15}$ in its phase I and a similar reach to Mu2e as final target.

%
\section{\bf CP asymmetry in $\mu \to 3e$}
\label{sec:Mu3e:AT}
%
Having outlined the list of observables in the muon sector, we now define observables that are indicators of possible CPV in LFV processes. Such quantities are always proportional to the polarization of the muon; unpolarized beams will wash out the angular dependence of the final states. Thankfully, present and future experiments deploy highly polarized high intensity muon beams, which makes the task of measuring CPV effects experimentally feasible. 

On the theory side, muon beams obtained from pions decaying at rest are expected to be 100\% polarized in the opposite direction of their momentum vector. Experimentally, the actual polarization at the muon stopping target is affected by depolarization effects and needs to be measured. For example, the average polarization measured by the MEG
experiment at PSI is 86\%~\cite{MEG:2015kvn} but higher polarization levels can be obtained in dedicated experiments, as the one performed on TRIUMF beam line, where the average measured polarization was $P_{\mu}=-0.99863\pm0.00088$~\cite{Jodidio:1986mz}. We assume in the following that a similar level of purity is attainable in forthcoming experiments.  

Considering $\mu^+\to e^+ e^- e^+$ decays with polarized muons, the angular distribution of the three $e^\pm$ tracks allows us to define a $T$-odd observable. Defining $\theta$ as the opening angle between the muon polarization and the direction of the outgoing electron, and $\phi$ as the azimuth angle of the muon polarization with respect to the plane of the decay~\cite{Kuno:1999jp, Bolton:2022lrg}, the time reversal acts on these angles as $T[\theta]=\theta$, $T[\phi]=2\pi-\phi$.

We can then define the $\at$ asymmetry as the normalized difference of number of events in the $\phi \in [-\pi,0]$ and $\phi\in[0,\pi]$ hemispheres:
\begin{equation}\label{eq:at-definition}
    \at\equiv \frac{\Gamma(\mu \to 3 e; c_\phi>0)-\Gamma(\mu \to 3 e; c_\phi<0)}{\Gamma(\mu\to 3 e)}  \,,
\end{equation}
where $c_\phi=\cos\phi$. The full expression of $A_T$ in terms of LEFT operators can be found in Refs.~\cite{Kuno:1999jp, Bolton:2022lrg}. To show the general behaviour, here we take the case where LFV dipoles and left-left and right-right four-leptons operators are non-zero. Under this assumption, we have 
\begin{widetext}
\begin{equation}\label{eq:at-estimate}
    \at \simeq -\frac{0.14P_\mu \left(\Im [c_{\ell\g}^{\mu e} (c_V^{RR})^*]+\Im [c_{\ell\g}^{e\mu} (c_V^{LL})^*]\right)}{|c_{\ell\g}^{\mu e}|^2+|c_{\ell\g}^{e\mu}|^2 + 8\times10^{-2} (|c_V^{RR}|^2+|c_V^{LL}|^2) + 0.2 (\Re [c_{\ell\g}^{\mu e} (c_V^{RR})^*]+\Re [c_{\ell\g}^{e\mu} (c_V^{LL})^*]) }\, ,
\end{equation}
\end{widetext}
where the numerical factors are to be intended as an indication of the real size of each contribution and $P_\mu$ is the magnitude of the muon polarization vector, expected to be $P_\mu\sim-1$ as discussed above. A similar equation can be obtained switching on left-right four-lepton operators.

From this structure we see immediately that a phase difference between dipole and four-lepton operators is needed to generate a non-vanishing $\at$. Furthermore, its allowed value is highly correlated with the bounds coming from ${\rm Br}(\mu\to e\gamma)$ and ${\rm Br}(\mu\to 3 e)$, which constraint the absolute values of the two LEFT coefficients. Finally, note that in the denominator of Eq.~\eqref{eq:at-estimate}, the dipole term is clearly the dominating term, justifying the approximation in Eq.~\eqref{eq:mu3e}. It follows that, for dipoles and four-lepton Wilson coefficients of ${\cal O}(1)$, the largest asymmetry achievable is of order of 10\%. Allowing larger hierarchies between $c_{\ell\g}^{\mu e}$ and $c_V^{RR}$, larger values of $A_T$ can be reached.

\paragraph*{\bf Other asymmetries:} 

The asymmetry in Eq.~\eqref{eq:at-definition} does not exhaust all the possible phase differences among the operators in Eq.~\eqref{eq:LEFT}. In particular we would like to construct observables sensitive to the CP violating phase among the flavor violating dipoles only, which in the basis of  Eq.~\eqref{eq:LEFT} can be written as $\Im[c_{\ell\g}^{\mu e}(c_{\ell\g}^{e\mu})^*]$.

A natural observable to look at is $\mu^+\to e^+ \gamma$, from which we would like to extract the CP violating parameter without relying on the $\mu^-\to e^- \gamma$ decay mode. As discussed in Refs.~\cite{Farzan:2007us, Vasquez:2015una}, one can define a CP-odd observable by measuring the azimuth angle $\phi_s$ between the spin of the outgoing positron and the plane spanned by the muon polarization and the momentum of the positron. The differential rate in the azimuth $\phi_s$ can be schematically written as 
\begin{eqnarray}\label{eq:diff-azimuth-meg}
\frac{d\Gamma(\mu\to e\gamma)}{d\phi_s}&\propto&P_\mu \Re[e^{i\phi_s} c_{\ell\g}^{\mu e}(c_{\ell\g}^{e\mu})^*]\sin\theta_s\ ,
\end{eqnarray}
where $\theta_s$ is the angle between the polarization of the positron and its
momentum and we have written only the parametric of the term sensitive to $\phi_s$ in the differential width. In order to define a CP-odd observable we then need a measurement of the spin of the electron and the azimuth angle, while the photon polarization is correlated to the electron spin after momentum and spin conservation in the two body decay are enforced~\cite{Vasquez:2015una}. 

Concretely, starting from Eq.~\eqref{eq:diff-azimuth-meg} we define
\begin{equation}
A_{\rm T}^{e\gamma}=P_\mu \frac{\Im[c_{\ell\g}^{\mu e}(c_{\ell\g}^{e\mu})^*]\sin\theta_s}{|c_{\ell\g}^{e\mu}|^2 + |c_{\ell\g}^{\mu e}|^2}\,.
\end{equation}
On top of the experimental challenge of measuring the electron spin, this asymmetry is also parametrically suppressed unless the polarization of the positron is selected to be perpendicular to its momentum. These two issues make this observable definitely more challenging to measure than the one defined in Eq.~\eqref{eq:at-definition}.\footnote{The precise definition is
\begin{equation}\label{eq:AT-meg}
    A_{\rm T}^{e\gamma}\equiv \frac{1}{\Gamma(\mu\to e \gamma)} \int _{0}^{2\pi}d\phi_s \frac{d\Gamma( \theta_s )}{d\phi_s}\mathrm{\mathrm{sign}(\cos\phi_s)}\ .
\end{equation}
}

The possibility of defining CP violating observables in $\mu\to e$ conversion was considered in Refs.~\cite{Davidson:2008ui,YaserAyazi:2008xzg,Vasquez:2015una}. In particular the same asymmetry as in Eq.~\eqref{eq:AT-meg} can be defined. The parametric dependence of the asymmetry is very similar to the one of $\mu\to e\gamma$, with the caveat that in $\mu\to e$ conversion experiments the muons have a smaller degree of polarization. This unavoidable depolarization effect makes it more difficult for this asymmetry to be observed experimentally. 

%
\section{A window onto the SMEFT}
\label{sec:Operators}
%

Having in mind an UV origin of the new physics effects in LFV muon decays, we need to compute the appropriate matching into SMEFT at the electroweak scale, $\mu_W$. The dimension-six lagrangian can be written as
\begin{equation}\label{eq:SMEFT}
    \mathscr{L}_{\rm SMEFT,6} = \sum_i \frac{\mathcal{C}_i}{\Lambda^2}\mathcal{Q}_i\,,
\end{equation}
where here $\Lambda$ is the same scale as in Eq.~\eqref{eq:LEFT}; the coefficents $\mathcal{C}_i$ are evaluated at this scale. The full basis of SMEFT operators can be found in Ref.~\cite{Grzadkowski:2010es}.

While the complete basis contains thousands of operators, we can restrict to a small set of them, listed in Table~\ref{tab:operator-set}, by using few considerations.  

Firstly, we require large CP violating effects dipole operators. This is easily realized by considering the non-hermitian dipole operators, $\Op_{eB}$ and $\Op_{eW}$. At the weak scale, these match directly at tree level into the dipole coefficients in Eq.~\eqref{eq:LEFT} as
\beq
c_{\ell\g}^{pr}(\mu_W) = \frac{v}{m_\mu}\left[ s_w {\cal C}_{eW}^{pr}(\mu_W) + c_w {\cal C}_{eB}^{pr}(\mu_W) \right]\,,
\eeq
where $p,r=1,2,3$ are generation indices, and $s_w$ and $c_w$ are the sine and cosine of the weak mixing angle, respectively. Note however, that in any UV realization we expect these dipoles to be generated at one-loop level. 

Secondly, we need to generate four-fermion structures at low energy to have a non-zero $\at$, as shown in Eq.~\eqref{eq:at-estimate}. This is fulfilled by including the four-lepton operators $\Op_{\ell\ell}$, $\Op_{ee}$ and $\Op_{\ell e}$, and lepton current $\times$ Higgs current operators, $\Op_{H\ell}^{(1),(3)}$ and $\Op_{He}$. These are highlighted in green and orange respectively in Table~\ref{tab:operator-set}. They give similar effects in the physics of $\mu\to 3 e$, as they match to the current-current operators in Eq.~\eqref{eq:LEFT} as a simple linear combination, $c_V \propto {\cal C}_{H\ell,He} + {\cal C}_{\ell\ell,ee,\ell e}$. 

For ease of the reader, we report in Appendix~\ref{app:matchings} the expressions of the LEFT coefficients in terms of SMEFT Wilson coefficients.\footnote{Both running and threshold corrections are taken into account in the matching done in Appendix~\ref{app:matchings}.} Given the matching described above and Eq.~\eqref{eq:at-estimate} it is easy to see that a large hierarchy between dipole and four-lepton operators is required in order to achieve $\at\sim\mathcal{O}(1)$. 

Under these circumstances the operators with the Higgs current are disfavoured to generate the required  four-leptons operators in the LEFT and for this reason are indicated in orange in Table~\ref{tab:operator-set}. The reason is that the presence of these operators at high energy generates operators with quark currents at low energy which directly contribute to $\mu\to e$ conversion processes~\cite{Davidson:2020hkf}. The stringent bounds on $\mu\to e$ conversion make it difficult for $\mu\to3e$ to be able to measure a large CP asymmetry in LFV decays.  

Lastly, we need to take into account other operators that generate or receive large mixing contributions from one-loop running and matching. Running from some high scale to $\mu_W$ via the RGE mixes operators; at leading-log, this mixing will result in a linear combination of Wilson coefficients evaluated at $\mu_W$.  Fermion-Higgs and four-fermion operators largely mix amongst each other by means of gauge interactions, ${\cal C}_{FH}\propto g_2^2 {\cal C}_{4F}$ (and viceversa), where $g_2$ is the $SU_L(2)$ gauge coupling. The same coupling will also control the leading-log terms in the matching with the LEFT in Eq.~\eqref{eq:LEFT}.

So far we have not introduced quarks in our discussion. Operators including heavy quarks, especially the top, can lead to large one-loop effects to mixing and matching. In particular, the operator $\Op_{lequ}$ will mix in the dipole as 
\beq\label{eq:running}
\Delta \mathcal{C}^{pr}_{eW,B} \propto \frac{y_t^2}{16\pi^2} \mathcal{C}^{pr33}_{\ell equ}\log\lp\frac{\Lambda}{\mu_W}\rp\,,
\eeq
where $y_t\sim1$ is the top quark Yukawa, and will contribute to the matching of $c_{\ell\g}$ with a similar coefficient. Similarly, the operators $\Op_{lq}^{(1)}$ and $\Op_{lq}^{(3)}$ will mix with the 4F and match into $c_V$. These operators are written in red highlight in Table~\ref{tab:operator-set}. 

Lastly, the most relevant finite contribution to dipoles comes from $\tau$ lepton threshold matching of $\Op_{\ell e}$ and reads
\beq\label{eq:thresholds}
\Delta c_{\ell\g}^{pr} \propto \frac{m_\tau/m_\mu}{16\pi^2}C_{\ell e}^{p33r}\,.
\eeq

\begin{table}[t!]
    \centering
    \caption{List of relevant SMEFT operators. We highlight in green the operators that generate the four-lepton operators in Eq.~\eqref{eq:LEFT}, while in red we show the relevant lepton-quark terms (see text for details).}
    \renewcommand{\arraystretch}{1.5} 
    \begin{tabular}{c|c}
    
        SMEFT name & structure\\
        \hline\hline
        $\mathcal{Q}_{eB}$ & $\bar{L}_r\sigma^{\mu\nu} E_s B_{\mu\nu}$ \\
        $\mathcal{Q}_{eW}$ & $\bar{L}_r\sigma^{\mu\nu}\tau^a E_s W^a_{\mu\nu}$  \\
         \hline
        \rowcolor{Grn!20}
        $\mathcal{Q}_{\ell\ell}$ & $\bar{L}_p \gamma^\mu L_r \bar{L}_s \gamma_\mu L_t$ \\
        \rowcolor{Grn!20}
        $\mathcal{Q}_{ee}$ & $\bar{E}_p \gamma^\mu E_r \bar{E}_s \gamma_\mu E_t$ \\
        \rowcolor{Grn!20}
        $\mathcal{Q}_{le}$ & $\bar{L}_p \gamma^\mu L_r \bar{E}_s \gamma_\mu E_t$ \\ 
           \hline
        \rowcolor{orange!20}
        $\mathcal{Q}^{(1)}_{H\ell}$ & $\bar{L}_r \gamma^\mu L_s \JH$ \\
        \rowcolor{orange!20}
        $\mathcal{Q}^{(3)}_{H\ell}$ & $\bar{L}_r \gamma^\mu \tau^a L_s \JHa$ \\
        \rowcolor{orange!20}
        $\mathcal{Q}_{He}$ & $\bar{E}_r \gamma^\mu E_s \JH$ \\
                    \hline
        \rowcolor{Red!20}$\mathcal{Q}^{(3)}_{\ell equ}$ &  $\epsilon_{ab}\bar {L}^a_p\sigma_{\mu\nu}E_r \bar{Q}^b_s \sigma^{\mu\nu}U_t$  \\
                        \hline
        \rowcolor{Red!20}$\mathcal{Q}_{\ell q}^{(1),(3)}$ &  ${\bar L}_p \g_\mu (\tau^I) L_r {\bar Q}_s \g_\mu (\tau^I) Q_t$  \\

    \end{tabular}
    \label{tab:operator-set}
\end{table}

At this point, the flavor structure of our problem restricts the possible avenues to the UV. If LFV dipoles are not generated at the scale $\Lambda$, these can only be induced sizeably by $\Op_{lequ}$ mixings. FH and 4F operators have to be generated at the UV scale as well, since no RGE effect can enhance them substantially compared to dipoles. The strong bound on the electron dipole moment leave as the only viable possibility to observe large CP violation effects in LFV decays heavy new physics with $\mu$-philic couplings. At the scale $\Lambda$ we need to have new particles mostly coupled to the second generation lepton and avoid coupling generating operators of the type (lepton current $\times$ Higgs current). In Sec.~\ref{sec:UVmodels} we exhibit a concrete framework realizing this coupling structure.

%
\section{Scanning}
\label{sec:plot}
%

With the ingredients detailed in the previous Sections, we can pinpoint the interesting region of parameter space.\footnote{We use the package {\tt DSixTools 2.0}~\cite{dsix,dsix2} to perform our numerical analysis, from RGE mixing to matching with the LEFT.} We fix the heavy scale to $\Lambda = 10^7$ GeV which is roughly the projected sensitivity on cutoff scale of ${\cal B}(\mu\to3e)$, assuming $\mathcal{O}(1)$ Wilson coefficient and flavor anarchy~\cite{Calibbi:2017uvl}. We then scan the parameter space defined by the relevant Wilson coefficients in SMEFT. These are imposed as initial conditions of the RGE at the scale $\Lambda$ and are motivated by the considerations of the previous section. After performing the full RGE running and matching we get the contributions of LEFT coefficients in Eq.~\eqref{eq:LEFT} to the various observables. 

In practice, we find it easier to reverse the flow of the RGE and run from the low scale to the cut-off scale $\Lambda$. Using built-in functions of {\tt DSixTools 2.0}, we get the full one-loop matching expression of the LEFT coefficients at the weak scale, $c(\mu_W) = \sum_i {\cal C}_i(\mu_W)$. We use here the approximation that QED running in the low energy EFT is negligible, see Sec.~\ref{sec:LEFT}, thus $c(\mu_W) = c(m_\mu, m_e)$. Given the set of ${\cal C}_i(\mu_W)$ in the linear combination, we can evolve them from $\mu_W$ to the UV cut-off scale. At the UV scale we then impose the desired initial conditions in order to obtain the low energy observables expressed as function of the initial Wilson coefficients in SMEFT. Specifically, we fix which coefficient is non-zero at $\Lambda$, their phase difference and relative size. We then scan over their absolute values to obtain our results.

\begin{figure*}[!t]
\centering
    \includegraphics[width=.45\linewidth]{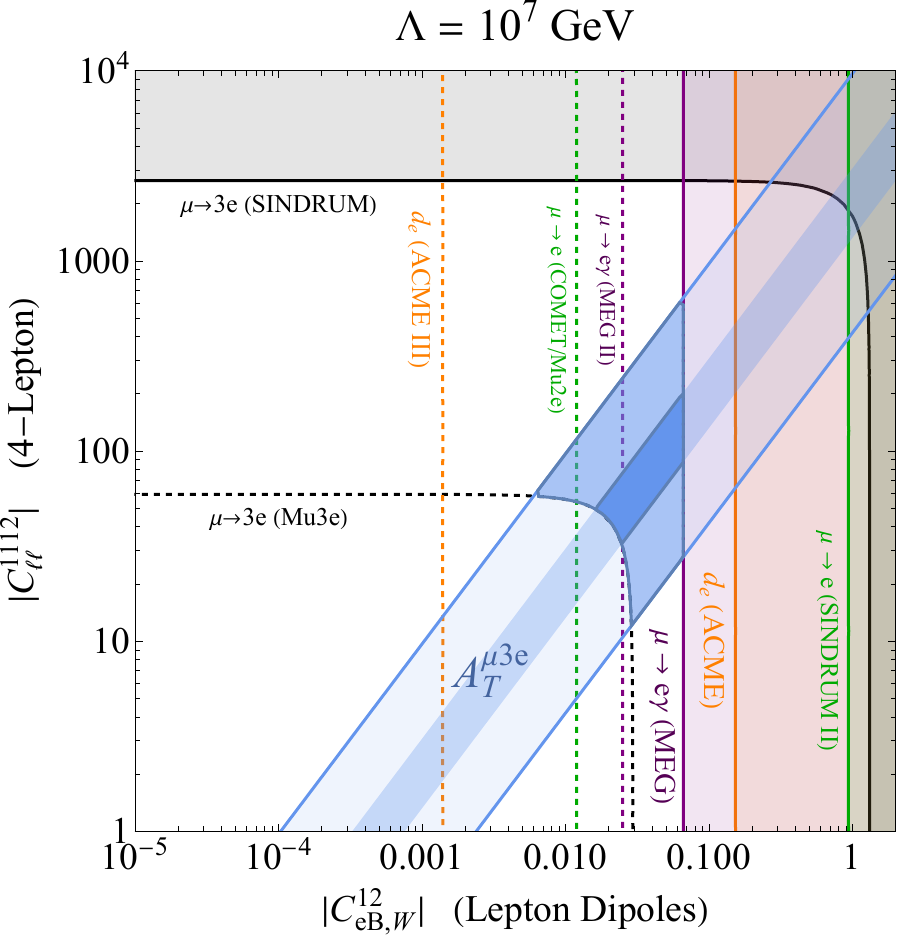}\hspace{1cm}
    \includegraphics[width=.438\linewidth]{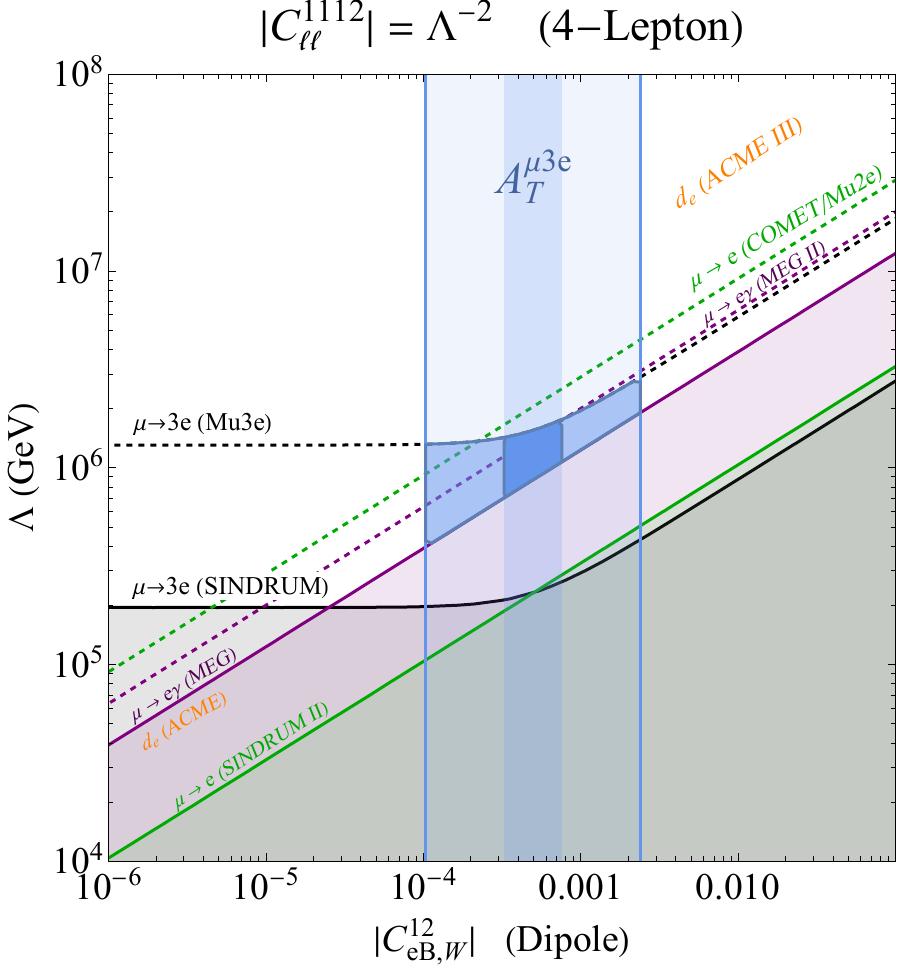}
    \caption{Limits in the Wilson coefficients parameter space of scenario 1, see text for details. \textbf{Left:} scan over Wilson coefficients, with fixed $\Lambda = 10^7$ GeV. \textbf{Right:} scan over dipole coefficients vs $\Lambda$, with the FH coefficient fixed to $1/\Lambda^2$. In both plots, the black, green, orange and purple shaded regions indicate bounds from $\mu\to3e$, $\mu\to e$, $d_e$ and $\mu\to e\g$ respectively. Dashed lines with the same color show projections on the respective observable from future experiments. The light (dark) blue region indicate where $A_T^{3e} > 0.1 (0.23)$; we further highlight the section probed by Mu3e.}
    \label{fig:LFV_Case_1}
\end{figure*}

In Fig.~\ref{fig:LFV_Case_1} we show our results for a reference scenario which will be easiliy mapped to the concrete model of Sec.~\ref{sec:UVmodels}. In the left plot we fix the UV scale to $\Lambda = 10^7$ GeV and we take as inputs the values
\beq\label{eq:inputs}
\begin{split}
&{\cal C}_{eB,W}^{12} = |x|e^{i\frac\pi2}\,,~~~ {\cal C}_{\ell\ell}^{1112} =  |y| \, ,\\ 
&{\cal C}_{eB,W}^{11} = |x|e^{i\frac\pi2}\times10^{-3}\, ,
\end{split}
\end{equation}
where the phase difference between dipole and four-lepton operator is maximal, and the flavor-diagonal dipole component is suppressed with respect to the off-diagonal. The latter requirement is necessary in order to generate small enough eEDM. The interesting parameter space in Fig.~\ref{fig:LFV_Case_1} features a sizeable hierarchy between four-lepton and dipole coefficients, as expected from Eq.~\eqref{eq:at-estimate}. The highlighted region indicate the part of parameter space where Mu3e phase-II will improve on the presently strongest bound ($\mu\to e\g$, purple line) and can observe ${\cal O}(1)$ CP violation in $\mu\to3e$ decays; for concreteness, we defined the latter as the regions where $\at > 0.1$ (ligther band) and $\at > 0.23$ (darker band). Complementary probes of LFV, as $\mu\to e\gamma$ and $\mu\to e$ conversion, will explore the same region in the near future; however, a positive signal from these won't be able to probe the CP nature of the process unless the final state polarization vectors will be measured. Similarly, future measurements of $d_e$ will cover the region of interest, despite assuming a largely suppressed diagonal term. A positive signal from the latter would be a smoking gun for CPV new physics.

In the right plot of Fig.~\ref{fig:LFV_Case_1} we show instead the scale probed by the LFV observables, by fixing the coefficient of the four-lepton operator to be 1 at $\Lambda = 10^7$ GeV, while keeping the other inputs as in Eq.~\eqref{eq:inputs}. We then leave the dipole coefficients and the scale $\Lambda$ as free parameters to scan over. Given the hierarchy discussed above, the required dipole coefficient is between ${\cal O}(10^{-3}-10^{-4})$, for a heavy scales of roughly $\sim10^6$ GeV.

%
\section{A possible origin: $\mu$-philic new fermions and heavy dark photon}
\label{sec:UVmodels}
%
Here we build an explicit model that realizes the setup described in the previous section. The first question to address is what type of new physics only modifies the lepton sectors as in Eq.~\eqref{eq:LEFT} at low energy. The UV sector needs to generate mostly the dipoles operators and the operators in green in Table \ref{tab:operator-set}. The other operator families (highlighted in orange and red in Table \ref{tab:operator-set}) give too large contributions to low-energy observables including quarks, as the $\mu\to e$ conversion rate and would make it impossible to observe CP violation in $\mu\to 3e$ decays. 

The new physics has to be $\mu$-specific to only select \eqref{eq:LEFT} at low energy. This can be realized adding new vector-like fermions with the quantum numbers of the SM leptons. The new sector needs also to be able to generate the dipoles with a minimal 'usage' of the Higgs, as not to introduced operators with the Higgs current. As such the Higgs field should not be too strongly coupled to the new fermions. The generation of the dipoles is then ascribed to the presence of a new massive vector, singlet under the SM, that we call dark photon, $A_\mu$. 

In formulas these conditions are encoded in the following sector 
\begin{equation}\label{eq:vectorlike}
\begin{split}
     \mathscr{L} &= m_E E \Psi_E +m_L L \Psi_L + M_E \Psi_E \bar\Psi_E + M_L \Psi_L \bar\Psi_L +\text{h.c} \\
     &-\frac14 A_{\mu\nu}^2 + \frac12 M_A^2 A _\mu A^\mu + \text{fermion kinetic terms} \\
     &+g_A A_\mu( c_L  \bar \Psi_L \gamma^\mu \Psi_L + c_E \bar \Psi_E \gamma^\mu \Psi_E)\,.
\end{split}
\end{equation}
This is the archetype of the new physics needed to generate Eq.~\eqref{eq:LEFT} (see Ref.~\cite{Contino:2006nn,Glioti:2024hye} for models with similar ingredients).

In the first line we have the mixing between the SM-like states and the new fermions, vector-like pairs with the quantum numbers of the SM lepton doublet ($\Psi_L$) and singlet ($\Psi_E$). The massive dark photon $A_\mu$ has coupling to such new fermions with strenght $g_A c_{L,E}$. Despite the fact that the new fermions have states with the quantum numbers of the SM neutrinos, no neutrino masses are generated.

In principle one can write other terms allowed by the symmetries. These are of two types: $i)$ possible renormalizable interactions including SM and new fermions with the Higgs; $ii)$ kinetic mixing between $A_\mu$ and the SM hypercharge. These will have the same effect in eq.~\eqref{eq:LEFT}, as they would generate 4-fermi operators with leptons and quarks. Our exercise shows that if CPV is searched for in $\mu\to 3e$, only a handful of operators need to be generated, then we set to zero these two types of interactions. The sector is therefore extremely weakly coupled to the Higgs and with a tiny kinetic mixing.

Since our muon observables are at very low energies, it will be enough to integrate out the new fermions and read out the effective field theory below their masses, the relevant scales being $M_{E,L}$ and $M_A$.

From the discussion of the previous section is clear that we need a hierarchical structure of $m_{L}$ and $m_{E}$. Namely the muon-philic case is realized with 
\begin{equation}
    (m_E)_{ij}=(m_E)_j \delta_{i2}\,,\quad (m_L)_{ij}=(m_L)_j \delta_{i2}\,,\quad
\end{equation}
that is, the new fermions couple exclusively with the second lepton generation. This can be thought as a limiting case. By resolving the mixing between SM and new fermions, one is led to redefine the SM states by a mixing in flavor space of an angle of order $\epsilon_{L,E}\sim m_{L,E}/M_{L,E}$. 
Once this mixing is resolved and the new SM eigenstates are identified, the flavor structure relevant for our discussion is schematically given by
\begin{equation}
\begin{split}
   &\sim Y\ell H^* \ell + Y \epsilon \Psi H^* \ell + M \Psi \bar\Psi\\
   &+ g_A A_\mu (\bar\Psi\gamma^\mu \Psi+  \epsilon \bar\ell \gamma^\mu \Psi +\epsilon^2 \bar\ell \gamma^\mu \ell)\ ,
   \end{split}
\end{equation}
where $Y$ is the SM lepton Yukawa matrix before electroweak symmetry breaking. The new physics can now be integrated out easily at tree- and loop-level.

Doing the matching at the high scale  we generate the following Wilson coefficients
\begin{equation}\label{eq:matching-model}
    \frac{\CC_{\ell\ell}}{\Lambda^2} \sim\frac{g_A^2}{M_A^2} \epsilon^4\,,\quad
  \frac{\CC_{eB,W}}{\Lambda^2} \sim\frac{g}{16\pi^2}\frac{g_A^2}{M_A^2} \epsilon \cdot \epsilon \cdot Y \cdot \epsilon\cdot \epsilon
\end{equation}
where we just highlighted the scaling with the flavor mixings $\epsilon_{ij}$ and the SM Yukawa can be written in the singular value decomposition as $Y=V_L Y_\ell^{\rm diag} V_R$. The Yukawa suppression together with the loop factor suppressed the dipole operators sufficiently compared to the four-leptons operators realizing naturallly the hierarchy highlighted in Fig~\ref{fig:LFV_Case_1}. Moreover it is easy to check that enough structure is present to generate a physical CP phase among the two sets of operators.

\section{Conclusions}
\label{sec:Conclusions}
In this work we explored the possibility of having large CP violation effects in the lepton flavor violating decay of the muon in three electrons, $\mu\to3e$. In Fig.~\ref{fig:LFV_Case_1} we show the parameter space where large CP violation is induced and will possibly be observable at Mu3e. Crucially, the diagonal entries of the dipole operators need to be largely suppressed with respect to the off-diagonal ones; otherwise, the eEDM imposes too stringent bounds on CP-violation in the lepton sector. Moreover, dimension 6 operators containing the SM Higgs or the SM quarks should be suppressed to avoid large new physics contributions to $\mu\to e$ conversion. The latter would constrain LFV too much to allow large CP violation to be observed at Mu3e.

All these constraints together select a very specific UV scenario where only a handful of SMEFT operators are generated (the white and green ones in Table~\ref{tab:operator-set}). We construct an explicit UV model that realizes this scenario, generating four-lepton operators at tree level and dipoles at one-loop with the right hierarchy among them in order to maximize their interference which controls the CP-violation. The new fermions of the model need to be heavier than $\sim10$ TeV to be consistent with the present bounds on the LFV. These new fermions can potentially lead to signatures at future colliders, e.g. at a high-energy muon collider.

Note that, apart from the eEDM $d_e$, no other present or future measurement of muonic LFV decays can probe CP asymmetries. From an optimistic view point, this poses Mu3e in the unique position of discovering both LFV and CP violation at the same time. An in-depth study of CP-violation effects in other LFV decay channels is warranted to extend our study (see for example Ref.~\cite{Kitano:2000fg} for interesting results on $\tau$ decays).  

\bigskip
\subsubsection*{Acknowledgements}
We thank Sasha Davidson and Marco Ardu for initial discussions that triggered this study. The work of DR is supported in part by the European Union - Next Generation EU through the PRIN2022 Grant n. 202289JEW4. MT and AT are thankful for the hospitality of CERN during the "Crossroads between Theory and Phenomenology" Workshop, where part of this work was carried out.

\bibliographystyle{h-physrev}
\bibliography{cedm}

\begin{thebibliography}{10}

\bibitem{ACME:2018yjb}
ACME, V.~Andreev {\em et~al.},
\newblock Nature {\bf 562}, 355 (2018).

\bibitem{Meisenhelder:2023qcq}
C.~Meisenhelder,
\newblock (2023).

\bibitem{Panico:2018hal}
G.~Panico, A.~Pomarol, and M.~Riembau,
\newblock JHEP {\bf 04}, 090 (2019), 1810.09413.

\bibitem{Cesarotti:2018huy}
C.~Cesarotti, Q.~Lu, Y.~Nakai, A.~Parikh, and M.~Reece,
\newblock JHEP {\bf 05}, 059 (2019), 1810.07736.

\bibitem{Mu3e:2020gyw}
Mu3e, K.~Arndt {\em et~al.},
\newblock Nucl. Instrum. Meth. A {\bf 1014}, 165679 (2021), 2009.11690.

\bibitem{Okada:1999zk}
Y.~Okada, K.-i. Okumura, and Y.~Shimizu,
\newblock Phys. Rev. D {\bf 61}, 094001 (2000), hep-ph/9906446.

\bibitem{Kuno:1999jp}
Y.~Kuno and Y.~Okada,
\newblock Rev. Mod. Phys. {\bf 73}, 151 (2001), hep-ph/9909265.

\bibitem{Bolton:2022lrg}
P.~D. Bolton and S.~T. Petcov,
\newblock Phys. Lett. B {\bf 833}, 137296 (2022), 2204.03468.

\bibitem{Muong-2:2008ebm}
Muon (g-2), G.~W. Bennett {\em et~al.},
\newblock Phys. Rev. D {\bf 80}, 052008 (2009), 0811.1207.

\bibitem{Adelmann:2021udj}
A.~Adelmann {\em et~al.},
\newblock (2021), 2102.08838.

\bibitem{MEG:2016leq}
MEG, A.~M. Baldini {\em et~al.},
\newblock Eur. Phys. J. C {\bf 76}, 434 (2016), 1605.05081.

\bibitem{MEGII:2023ltw}
MEG II, K.~Afanaciev {\em et~al.},
\newblock Eur. Phys. J. C {\bf 84}, 216 (2024), 2310.12614.

\bibitem{MEGII:2018kmf}
MEG II, A.~M. Baldini {\em et~al.},
\newblock Eur. Phys. J. C {\bf 78}, 380 (2018), 1801.04688.

\bibitem{BELLGARDT19881}
U.~Bellgardt {\em et~al.},
\newblock Nuclear Physics B {\bf 299}, 1 (1988).

\bibitem{Blondel:2013ia}
A.~Blondel {\em et~al.},
\newblock (2013), 1301.6113.

\bibitem{SINDRUMII:2006dvw}
SINDRUM II, W.~H. Bertl {\em et~al.},
\newblock Eur. Phys. J. C {\bf 47}, 337 (2006).

\bibitem{Mu2e:2022ggl}
Mu2e, F.~Abdi {\em et~al.},
\newblock Universe {\bf 9}, 54 (2023), 2210.11380.

\bibitem{COMET:2018auw}
COMET, R.~Abramishvili {\em et~al.},
\newblock PTEP {\bf 2020}, 033C01 (2020), 1812.09018.

\bibitem{Calibbi:2017uvl}
L.~Calibbi and G.~Signorelli,
\newblock Riv. Nuovo Cim. {\bf 41}, 71 (2018), 1709.00294.

\bibitem{Jenkins:2017jig}
E.~E. Jenkins, A.~V. Manohar, and P.~Stoffer,
\newblock JHEP {\bf 03}, 016 (2018), 1709.04486.

\bibitem{Dekens_2019}
W.~Dekens and P.~Stoffer,
\newblock Journal of High Energy Physics {\bf 2019} (2019).

\bibitem{Kitano:2002mt}
R.~Kitano, M.~Koike, and Y.~Okada,
\newblock Phys. Rev. D {\bf 66}, 096002 (2002), hep-ph/0203110,
\newblock [Erratum: Phys.Rev.D 76, 059902 (2007)].

\bibitem{Haxton:2024lyc}
W.~Haxton, K.~McElvain, T.~Menzo, E.~Rule, and J.~Zupan,
\newblock (2024), 2406.13818.

\bibitem{Wintz:1998rp}
P.~Wintz,
\newblock Conf. Proc. C {\bf 980420}, 534 (1998).

\bibitem{Mu2e:2014fns}
Mu2e, L.~Bartoszek {\em et~al.},
\newblock (2014), 1501.05241.

\bibitem{MEG:2015kvn}
MEG, A.~M. Baldini {\em et~al.},
\newblock Eur. Phys. J. C {\bf 76}, 223 (2016), 1510.04743.

\bibitem{Jodidio:1986mz}
A.~Jodidio {\em et~al.},
\newblock Phys. Rev. D {\bf 34}, 1967 (1986),
\newblock [Erratum: Phys.Rev.D 37, 237 (1988)].

\bibitem{Farzan:2007us}
Y.~Farzan,
\newblock JHEP {\bf 07}, 054 (2007), hep-ph/0701106.

\bibitem{Vasquez:2015una}
J.~C. Vasquez,
\newblock JHEP {\bf 09}, 131 (2015), 1504.05220.

\bibitem{Davidson:2008ui}
S.~Davidson,
\newblock (2008), 0809.0263.

\bibitem{YaserAyazi:2008xzg}
S.~Yaser~Ayazi and Y.~Farzan,
\newblock JHEP {\bf 01}, 022 (2009), 0810.4233.

\bibitem{Grzadkowski:2010es}
B.~Grzadkowski, M.~Iskrzynski, M.~Misiak, and J.~Rosiek,
\newblock JHEP {\bf 10}, 085 (2010), 1008.4884.

\bibitem{Davidson:2020hkf}
S.~Davidson,
\newblock JHEP {\bf 02}, 172 (2021), 2010.00317.

\bibitem{dsix}
A.~Celis, J.~Fuentes-Martin, A.~Vicente, and J.~Virto,
\newblock Eur. Phys. J. C {\bf 77}, 405 (2017), 1704.04504.

\bibitem{dsix2}
J.~Fuentes-Martin, P.~Ruiz-Femenia, A.~Vicente, and J.~Virto,
\newblock Eur. Phys. J. C {\bf 81}, 167 (2021), 2010.16341.

\bibitem{Contino:2006nn}
R.~Contino, T.~Kramer, M.~Son, and R.~Sundrum,
\newblock JHEP {\bf 05}, 074 (2007), hep-ph/0612180.

\bibitem{Glioti:2024hye}
A.~Glioti, R.~Rattazzi, L.~Ricci, and L.~Vecchi,
\newblock (2024), 2402.09503.

\bibitem{Kitano:2000fg}
R.~Kitano and Y.~Okada,
\newblock Phys. Rev. D {\bf 63}, 113003 (2001), hep-ph/0012040.

\end{thebibliography}

\clearpage
\onecolumngrid

\begin{appendix}

\section{SMEFT - LEFT Matching}
\label{app:matchings}

Here we provide the expression of the LEFT coefficients in eq.~\eqref{eq:LEFT}, evaluated at the matching scale, $\mu_W$, as function of the relevant SMEFT coefficients, Table~\ref{tab:operator-set}, evaluated at the heavy scale $\Lambda$. We obtain these results making use of the functions available in {\tt DSixTools 2.0}. Additionally, we neglect numerically small contributions, and retain only the most significant terms.
\beq
\begin{split}
    c_{\ell\g}^{pr}(\mu_W) &= 10^3  \lp 1.4 ~ \CC_{eB}^{pr} - 0.8 ~ \CC_{eW}^{pr} \rp + 0.1 ~ \CC_{\ell e}^{p33r} +  \lp 7.9 ~ \CC_{eB}^{pr} - 4.4 ~ \CC_{eW}^{pr} + 0.1~ \CC_{\ell equ}^{(3),pr23} \rp \log\lp\frac{\Lambda}{\mu_W}\rp \,, \\
    c_V^{LL} (\mu_W) &= \CC_{\ell\ell}^{1112} - 0.3 ~ \lp \CC_{H\ell}^{(1),12} + \CC_{H\ell}^{(3),12} \rp \\
    &- 10^{-4}  \lp 4.5 ~ \CC_{H\ell}^{(1),12} + 6.1 ~ \CC_{H\ell}^{(3),12} + 4.5 ~ \CC_{\ell q}^{(1),1233} - 6.3 ~ \CC_{\ell q}^{(3),1233} - 2.9 ~ \CC_{\ell u}^{1233} \rp \log\lp\frac{\Lambda}{\mu_W}\rp \,, \\
    c_V^{RR} (\mu_W) &=  \CC_{ee}^{1112} + 0.2 ~ \CC_{He}^{12} - 10^{-4}\lp 4.2 ~ \CC_{eu}^{1233} - 2.4 ~ \CC_{He}^{12} - 2.9 ~ \CC_{qe}^{3312}  \rp \log\lp\frac{\Lambda}{\mu_W}\rp \,, \\
    c_V^{LR} (\mu_W) &= \CC_{\ell e}^{1211} + 0.5 ~ \lp \CC_{H\ell}^{(1),12} + \CC_{H\ell}^{(3),12} \rp \\
    &+ 10^{-4}  \lp 4.9 ~ \CC_{H\ell}^{(1),12} + 7.2 ~ \CC_{H\ell}^{(3),12} + 6.7 ~ \CC_{\ell q}^{(1),1233} - 9.4 ~ \CC_{\ell q}^{(3),1233} - 7.4 ~ \CC_{\ell u}^{1233} \rp \log\lp\frac{\Lambda}{\mu_W}\rp \,, \\
    c_V^{RL} (\mu_W) &= \CC_{ee}^{1112} - 0.5 ~ \CC_{He}^{12} + 10^{-4}\lp 6.8 ~ \CC_{eu}^{1233} - 9.0 ~ \CC_{He}^{12} - 7.8 ~ \CC_{qe}^{3312}  \rp \log\lp\frac{\Lambda}{\mu_W}\rp \,,
\end{split}
\eeq
where all SMEFT coefficients are evaluated at the high scale, $\CC\equiv\CC(\Lambda)$.

\end{appendix}

\end{document}